\documentclass[journal]{IEEEtran}

\usepackage[utf8]{inputenc}
\usepackage{amsmath,amssymb,amsfonts}
\usepackage{algorithm}
\usepackage{algpseudocode}
\usepackage{graphicx}
\usepackage{textcomp}
\usepackage{xcolor}
\usepackage{cite}
\usepackage{hyperref}
\usepackage{booktabs}

\usepackage{algpseudocode}

\usepackage{amsmath}
\usepackage{chngcntr}
\counterwithout{equation}{section}

\begin{document}

% Başlık Ayarı
\title{Why Model Selection Fails in Time Series Forecasting: An Empirical Study of Instability Across Data Regimes}

%%%%%%%%%%%%%%%%%%%
\author{
\vspace{0.5em}
Tahir Cetin Akinci, and Alfredo A. Martinez-Morales
\vspace{0.8em} \\

\small University of California Riverside, Bourns College of Engineering, Ce-Cert, Riverside, California, USA
\vspace{0.5em} \\

\small \textit{Emails: tahircetin.akinci@ucr.edu}
}
%%%%%%%%%%%%%%%%%%%

\maketitle

\begin{abstract}
Time series forecasting models often exhibit inconsistent performance across datasets with varying statistical and structural properties. Despite the wide range of available forecasting techniques, it remains unclear whether model selection can be reliably guided by simple data characteristics. This paper investigates why rule-based model selection fails in time series forecasting by analyzing the relationship between data-regime descriptors and model performance. A descriptor-based framework is introduced to characterize time series using measurable properties, including trend strength, seasonality, noise level, and temporal dependence. Based on these descriptors, a rule-based selection mechanism is formulated to map data regimes to candidate forecasting models. The approach is evaluated on multiple real-world datasets across different domains and forecasting horizons. The results show that rule-based model selection achieves low accuracy, with correct model identification occurring in only a small fraction of cases. Significant discrepancies are observed between recommended and empirically optimal models, particularly in noisy and mixed regimes. Further analysis reveals that model performance is highly sensitive to both dataset characteristics and forecasting horizon, resulting in substantial ranking instability across scenarios. These findings explain why simple heuristic rules fail to generalize and demonstrate that forecasting performance cannot be reliably predicted using static, descriptor-based approaches. This study provides empirical evidence that model selection in time series forecasting is inherently context-dependent and highlights the need for more adaptive, data-driven strategies.
\end{abstract}

%%%%%%%%%%%%%%%%%%%%%%%%%%%%%%%%%%%%%%%%%%%%%%%%%%%%%%%%%%%%%%
%%%%%%%%%%%%%%%%%%%%%%%%%%%%%%%%%%%%%%%%%%%%%%%%%%%%%%%%%%%%%%

%%%%%%%%%%%%%%%%%%%%%%%%% SECTION 1 %%%%%%%%%%%%%%%%%%%%%%%%%%%%%%%%%
\section{Introduction}
%%%%%%%%%%%%%%%%%%%%%%%%%%%%%%%%%%%%%%%%%%%%%%%%%%%%%%%%%%%%%

Time series forecasting is a core component of modern data-driven systems, supporting decision-making in domains such as energy planning, financial analysis, transportation, and industrial monitoring \cite{saravana2026}. As forecasting workflows increasingly rely on training multiple models and evaluating them through automated pipelines, the reliability and generalization of forecasting models have become critical challenges \cite{hassler2026}. However, despite substantial progress in forecasting methodologies-ranging from classical statistical approaches such as ARIMA and exponential smoothing to machine learning and deep learning models-their performance remains highly sensitive to the characteristics of the underlying data \cite{makridakis2018m4, hyndman2018forecasting}.

In practice, it is frequently observed that a model performs well on one dataset but fails on another, even when the datasets appear similar at first glance. This variability is not incidental, but instead reflects a mismatch between the assumptions embedded in the model and the statistical stcan captureecasting models inherently rely on assumptions such as stationarity, smoothness, linearity, and temporal dependency structures \cite{mohammed2025}. When these assumptions align with the data, models tend to perform reliably; when they do not, performance deteriorates. From a theoretical perspective, this behavior can be interpreted through concepts such as inductive bias, model capacity, and the bias--variance tradeoff. These principles help explain why models with strong structural assumptions perform well under stable and low-complexity regimes but fail under non-stationary or structurally complex conditions, where statistical properties evolve over time \cite{hamilton1994time}.

Existing research has predominantly focused on developing new models or improving predictive performance through architectural innovations and optimization strategies \cite{krishnan2025}. While these efforts have led to significant advances, they provide limited guidance on how to select appropriate models for new datasets. In many real-world applications, practitioners still rely on trial-and-error strategies, where multiple models are trained and evaluated retrospectively \cite{knapen2026}. This approach is computationally expensive, difficult to reproduce, and often yields inconsistent results, particularly when datasets exhibit non-stationarity, structural breaks, intermittency, or complex seasonal patterns.

A fundamental limitation in the literature is the lack of a systematic framework that connects data characteristics to model behavior. To the best of our knowledge, limited work systematically establishes a unified and interpretable link between data properties and model performance. Time series data are inherently heterogeneous, with varying levels of trend, seasonality, noise, sparsity, and regime shifts \cite{hyndman2018forecasting}. At the same time, forecasting models differ in their inductive biases, which determine the types of patterns they are capable of capturing. For example, linear models are often effective for stable and low-noise signals, while more flexible architectures, such as recurrent or attention-based models, are better suited for capturing nonlinear dependencies and long-range interactions \cite{bandara2020lstm}. However, these relationships are rarely formalized in a way that can guide model selection in practice, particularly in large-scale empirical forecasting settings \cite{makridakis2018m4}.

To address this gap, this paper investigates why rule-based model selection fails in time series forecasting. Rather than assuming that statistical descriptors can reliably determine the optimal model, the proposed framework is used as an analytical tool to examine the relationship between measurable data characteristics and model behavior. The approach defines a set of quantitative descriptors capturing key statistical and structural properties of time series data, including trend strength, seasonal structure, noise level, non-stationarity, intermittency, and structural breaks \cite{hyndman2018forecasting, hamilton1994time}. These descriptors are used to represent data regimes, which are then analyzed in relation to the empirical performance of different model families.

The framework integrates statistical diagnostics with model assumption profiling and operationalizes their interaction through structured selection rules, enabling a systematic evaluation of whether such mappings can generalize across datasets and forecasting horizons. This problem is closely related to broader challenges in empirical forecasting, where model performance is often inconsistent across datasets and evaluation settings \cite{makridakis2018m4, makridakis2020m5}. 

Unlike prior approaches, the proposed formulation attempts to unify statistical diagnostics and model assumption characteristics within a single operational selection mechanism. While similar ideas have been explored in machine learning-based forecasting and meta-learning approaches, these relationships are rarely formalized in a transparent and interpretable manner \cite{bandara2020lstm}. This perspective allows the framework to be used as a baseline for analyzing the limitations of rule-based selection strategies in heterogeneous forecasting environments.

The proposed approach is evaluated using a diverse set of real-world datasets spanning multiple domains, including energy systems, financial signals, and transportation data. The empirical results indicate that the relationship between data-regime descriptors and model performance is more complex than commonly assumed. While certain descriptors provide partial insight into model suitability, the overall effectiveness of rule-based model selection remains limited. In particular, discrepancies between recommended and empirically optimal models are observed across several dataset-horizon cases, and model rankings exhibit substantial variability depending on both data characteristics and forecasting horizon \cite{makridakis2018m4, makridakis2020m5, bergmeir2018note}. 

These findings suggest that regime-aware descriptors alone are insufficient to ensure reliable model selection and are consistent with broader evidence that forecasting performance is highly context-dependent and difficult to generalize across heterogeneous datasets \cite{hyndman2018forecasting}.

This paper makes the following key contributions:
\begin{itemize}
    \item We introduce a data-regime-based representation of time series that captures essential statistical and structural properties relevant to forecasting.
    \item We develop a structured framework that enables systematic analysis of the relationship between data characteristics and model behavior.
    \item We provide empirical evidence highlighting the limitations of rule-based model selection across heterogeneous datasets and forecasting horizons.
    \item We demonstrate that model performance is highly context-dependent, influenced jointly by data regimes and forecasting horizon, leading to inherent selection instability.
\end{itemize}

Overall, this study reframes model selection as a context-dependent problem and provides empirical insights into why simple, static selection rules fail to generalize across diverse time series scenarios. :contentReference[oaicite:0]{index=0}

%%%%%%%%%%%%%%%%%%%%%%%%%%%%%%%%%%%%%%%%%%%%%%%%%%%%%%%%%%%%%%%%%%
%%%%%%%%%%%%%%%%%%%%%%% SECTION 2 %%%%%%%%%%%%%%%%%%%%%%%%%%%%%%%%
\section{Data-Regime Characterization and Framework Formulation}
%%%%%%%%%%%%%%%%%%%%%%%%%%%%%%%%%%%%%%%%%%%%%%%%%%%%%%%%%%%%%%%%%%

This section formalizes the proposed data-regime-based framework and defines a mapping from measurable data characteristics to model suitability.

%%%%%%%%%%%%%%%%%%%%%%%%%%%%%%%%%%%%%%%%%%%
\subsection{Time Series Representation}
%%%%%%%%%%%%%%%%%%%%%%%%%%%%%%%%%%%%%%%%%%%

Let $x(t)$ denote a univariate time series of length $T$:
\begin{equation}
x(t) = s(t) + n(t), \quad t = 1,2,\dots,T
\label{eq:signal}
\end{equation}
where $s(t)$ captures structured dynamics and $n(t)$ represents stochastic noise. Equation (\ref{eq:signal}) motivates extracting descriptors that quantify both structure and uncertainty \cite{rao1976}.

%%%%%%%%%%%%%%%%%%%%%%%%%%%%%%%%%%%%%%%%%%%
\subsection{Data Regime Descriptor Vector}
%%%%%%%%%%%%%%%%%%%%%%%%%%%%%%%%%%%%%%%%%%%

Each series is mapped to a descriptor vector:
\begin{equation}
\mathbf{r} = [r_1, r_2, \dots, r_d]
\label{eq:descriptor_vector}
\end{equation}
which summarizes the statistical regime.

These descriptors are selected because they capture the dominant sources of variation in univariate time series, including trend, periodicity, stochastic variability, structural change, and sparsity. Together, they provide a compact yet expressive representation of the data regime while maintaining computational tractability. Alternative descriptors exist, but the chosen set balances interpretability, generality, and efficiency across diverse datasets. Descriptor stability under noise, missing data, and short time series is further discussed in Section V.

%%%%%%%%%%%%%%%%%%%%%%%%%%%%%%%%%%%%%%%%%%%%
\begin{itemize}
    \item \textbf{Trend Strength ($r_1$)}:
    \begin{equation}
    r_1 = \frac{|\hat{\beta}|}{\mathrm{Var}(x)}
    \label{eq:trend}
    \end{equation}
    where $\hat{\beta}$ is the slope of a linear fit. Equation (\ref{eq:trend}) normalizes trend magnitude by signal variability.

    \item \textbf{Seasonality ($r_2$)}:
    \begin{equation}
    r_2 = \frac{\max_{f \in \mathcal{F}} P(f)}{\sum_{f} P(f)}
    \label{eq:seasonality}
    \end{equation}
    where $P(f)$ is spectral power. Equation (\ref{eq:seasonality}) captures dominant periodicity via spectral concentration.

    \item \textbf{Noise Level ($r_3$)}:
    \begin{equation}
    r_3 = \frac{\sigma_n^2}{\sigma_x^2}
    \label{eq:noise}
    \end{equation}
    where $\sigma_n^2$ is residual variance. Equation (\ref{eq:noise}) measures the proportion of noise.

    \item \textbf{Non-Stationarity ($r_4$)}:
    \begin{equation}
    r_4 = 1 - p_{\text{ADF}}
    \label{eq:nonstationary}
    \end{equation}
    where $p_{\text{ADF}}$ is the ADF p-value. Equation (\ref{eq:nonstationary}) increases with non-stationarity.

    \item \textbf{Intermittency ($r_5$)}:
    \begin{equation}
    r_5 = \frac{1}{T} \sum_{t=1}^{T} \mathbb{I}(|x(t)| < \epsilon)
    \label{eq:intermittency}
    \end{equation}
    where $\mathbb{I}$ is the indicator. Equation (\ref{eq:intermittency}) quantifies sparsity.

    \item \textbf{Structural Breaks ($r_6$)}:
    \begin{equation}
    r_6 = \frac{N_{\text{cp}}}{T}
    \label{eq:breaks}
    \end{equation}
    where $N_{\text{cp}}$ is the number of change points. Equation (\ref{eq:breaks}) reflects regime shifts.
\end{itemize}

Thus,
\begin{equation}
\mathbf{r} = f(x(t))
\label{eq:feature_map}
\end{equation}
where $f(\cdot)$ aggregates Eqs. (\ref{eq:trend})--(\ref{eq:breaks}) \cite{wang2006, priestley1981, said1984}. All descriptors are min--max normalized using statistics computed from the training set to ensure comparable scaling and consistent weighting across components.

%%%%%%%%%%%%%%%%%%%%%%%%%%%%%%%%%%%%%%%%%%
\subsection{Model Compatibility Function}
%%%%%%%%%%%%%%%%%%%%%%%%%%%%%%%%%%%%%%%%%%

Let $\mathcal{M} = \{M_1,\dots,M_K\}$ be candidate models. A compatibility score is defined as:

%%%%%%%%%%%%%%%%%%%%%%%%%
\begin{equation}
\phi_k(\mathbf{r}) = \mathbf{w}_k^T \mathbf{r}
\label{eq:compatibility_linear}
\end{equation}
%%%%%%%%%%%%%%%%%%%%%%%%%

Equation (\ref{eq:compatibility_linear}) provides a linear scoring rule suitable for interpretable, rule-based selection. In the learned mapping variant, the weights $\mathbf{w}_k$ are estimated using historical forecasting tasks where model performance labels are available, enabling supervised or meta-learning-based calibration of compatibility scores.

Alternative formulations include:
\begin{equation}
\phi_k(\mathbf{r}) = \sigma(\mathbf{w}_k^T \mathbf{r})
\label{eq:compatibility_logistic}
\end{equation}
\begin{equation}
\phi_k(\mathbf{r}) = -\|\mathbf{r} - \boldsymbol{\mu}_k\|_2
\label{eq:compatibility_distance}
\end{equation}

Equation (\ref{eq:compatibility_logistic}) enables probabilistic interpretation and is suitable when model selection is framed as a classification task. Equation (\ref{eq:compatibility_distance}) measures similarity to a prototypical regime $\boldsymbol{\mu}_k$, which can be estimated from historical datasets where model $M_k$ performs consistently well \cite{talagala2018, mohshini2024}.

%%%%%%%%%%%%%%%%%%%%%%%%%%%%%%%%%%%%%%%%%%%%%%%%%%%%
\subsection{Regime--Model Mapping}
%%%%%%%%%%%%%%%%%%%%%%%%%%%%%%%%%%%%%%%%%%%%%%%%%%%%

Model selection is defined as:
\begin{equation}
M^* = \arg\max_{M_k \in \mathcal{M}} \phi_k(\mathbf{r})
\label{eq:model_selection}
\end{equation}

Equation (\ref{eq:model_selection}) selects the model with highest compatibility. The mapping can be implemented as:
\begin{itemize}
    \item Rule-based selection using expert-defined thresholds
    \item Learned mapping using meta-learning, few-shot learning, or surrogate models trained on prior tasks
\end{itemize}

This enables adaptation to both expert-driven and data-driven forecasting environments.

%%%%%%%%%%%%%%%%%%%%%%%%%%%%%%%%%%%%%%%%%%%%%%%%%%%%
\subsection{Algorithmic Implementation}
%%%%%%%%%%%%%%%%%%%%%%%%%%%%%%%%%%%%%%%%%%%%%%%%%%%%

While Eq.~(\ref{eq:model_selection}) defines the model selection process in a mathematical form, the practical implementation can be expressed as a rule-based selection algorithm. The procedure is summarized in Algorithm~\ref{alg:selection}, which operationalizes the mapping from descriptor space to model choice.

%%%%%%%%%%%%%%%%%%%%%%%%%%%%%%%%%%%%%%%%%%%%%%%%%%%%
\begin{algorithm}[htbp]
\caption{Regime-Based Model Selection Algorithm}
\label{alg:selection}
\begin{algorithmic}[1]
\Require Time series $x(t)$
\Ensure Selected forecasting model $M^*$

\State Compute descriptor vector $\mathbf{r}$ using Eq.~(\ref{eq:feature_map})

\If{$r_3$ is low and autocorrelation is high}
    \State $M^* \leftarrow$ ARIMA
\ElsIf{$r_3$ is high}
    \State $M^* \leftarrow$ Random Forest
\ElsIf{$r_2$ is high}
    \State $M^* \leftarrow$ ETS
\Else
    \State $M^* \leftarrow$ Naive
\EndIf

\Return $M^*$
\end{algorithmic}
\end{algorithm}
%%%%%%%%%%%%%%%%%%%%%%%%%%%%%%%%%%%%%%%%%%%%%%%%%%%%

\noindent The algorithm provides a deterministic mapping between data-regime 
descriptors and candidate forecasting models, serving as an interpretable 
baseline for evaluating the effectiveness of rule-based model selection 
strategies. The rule set in Algorithm~\ref{alg:selection} is intentionally 
simple and reflects commonly used practitioner heuristics (e.g., low noise 
$\rightarrow$ ARIMA, high noise $\rightarrow$ RF, strong seasonality 
$\rightarrow$ ETS). The purpose of this formulation is not to design an 
optimal selector, but rather to provide a transparent and easily analyzable 
mechanism for understanding why static, descriptor-driven model selection 
fails to generalize across heterogeneous datasets and forecasting horizons. 
This simplified structure enables the instability and limitations of 
rule-based selection to be examined in a controlled and interpretable manner.

%%%%%%%%%%%%%%%%%%%%%%%%%%%%%%%%%%%%%%%%%%%%%%%%%%%%
\subsection{Interpretability}
%%%%%%%%%%%%%%%%%%%%%%%%%%%%%%%%%%%%%%%%%%%%%%%%%%%%

To better understand how data-regime descriptors influence model selection, we introduce a local sensitivity-based interpretability measure. Specifically, the influence of each descriptor is quantified through the gradient of the scoring function with respect to the descriptor vector:

%%%%%%%%%%%%%% EQ 14 %%%%%%%%%%%%%%%%%
\begin{equation}
I_i = \frac{\partial \phi_k(\mathbf{r})}{\partial r_i}
\label{eq:interpretability}
\end{equation}
%%%%%%%%%%%%%%%%%%%%%%%%%%%%%%%%%%%%%%

Equation (\ref{eq:interpretability}) measures the sensitivity of the model compatibility score to changes in descriptor $r_i$, providing a first-order approximation of descriptor importance. A positive value of $I_i$ indicates that increasing $r_i$ improves the compatibility of model $k$, whereas a negative value suggests that the descriptor negatively affects model suitability. 

This formulation enables a localized interpretation of the decision mechanism by capturing how small perturbations in the descriptor space influence model selection outcomes. Such gradient-based interpretations are consistent with established approaches for analyzing nonlinear decision functions, where model outputs depend on complex interactions among input variables \cite{montavon2017}.

%%%%%%%%%%%%%%%%%%%% G %%%%%%%%%%%%%%%%%%%%%%%%%%%%
\subsection{Computational Complexity}
%%%%%%%%%%%%%%%%%%%%%%%%%%%%%%%%%%%%%%%%%%%%%%%%%%%%

The computational cost of the framework is dominated by descriptor extraction and model selection. Descriptor extraction involves operations such as spectral analysis and change-point detection, leading to an overall complexity of $\mathcal{O}(T \log T)$. The model selection step requires evaluating compatibility scores across $K$ models and $d$ descriptors, resulting in a complexity of $\mathcal{O}(Kd)$ \cite{faloutsos1997, lin2007}.

%%%%%%%%%%%%%%%%%%%%%%%%%%%%%%%%%%%%%%%%%%%%%%%%%%%%
\subsection{Framework Summary}
%%%%%%%%%%%%%%%%%%%%%%%%%%%%%%%%%%%%%%%%%%%%%%%%%%%%

The workflow is:
\begin{enumerate}
    \item Extract descriptors $\mathbf{r}$ using Eq. (\ref{eq:feature_map})
    \item Compute $\phi_k(\mathbf{r})$ via Eqs. (\ref{eq:compatibility_linear})--(\ref{eq:compatibility_distance})
    \item Select $M^*$ using Eq. (\ref{eq:model_selection})
    \item Interpret using Eq. (\ref{eq:interpretability})
\end{enumerate}

This yields a structured, interpretable, and reproducible model selection process.

%%%%%%%%%%%%%%%%%%%%%%%%%%%%%%%%%%%%%%%%%%%%%%%%%%%%%%%%%%%%%
%%%%%%%%%%%%%%%%%%%%% SECTION 3 %%%%%%%%%%%%%%%%%%%%%%%%%%%%%
\section{Experimental Design and Validation}
%%%%%%%%%%%%%%%%%%%%%%%%%%%%%%%%%%%%%%%%%%%%%%%%%%%%%%%%%%%%%

This section evaluates the proposed data-regime-based model selection framework in terms of forecasting accuracy, robustness, and computational efficiency.

%%%%%%%%%%%%%%%%%%%%%%%%%%%%%%%%%%%%%%%%%%%%%%%%%%%%
\subsection{Dataset Collection}
%%%%%%%%%%%%%%%%%%%%%%%%%%%%%%%%%%%%%%%%%%%%%%%%%%%%

Experiments are conducted on four real-world datasets obtained from publicly available benchmark sources. 
The Tourism dataset corresponds to the well-known tourism demand dataset introduced by Hyndman et al. \cite{hyndman2018forecasting}. 
The Energy dataset is derived from the M4 forecasting competition dataset \cite{makridakis2018m4}, which consists of a large collection of time series from multiple domains and frequencies. 
The Electricity dataset corresponds to the Electricity Load Diagrams 2011--2014 dataset obtained from the UCI Machine Learning Repository \cite{uci_electricity}, containing consumption measurements from 370 clients recorded at 15-minute intervals. 
Finally, the PGCB dataset is obtained from the UCI Machine Learning Repository \cite{uci_pgcb}, providing hourly generation and demand data from the Bangladesh national grid. 
These datasets are selected to ensure diversity in statistical properties and to evaluate model performance across different application domains.

All datasets are preprocessed using a consistent pipeline. Missing values are imputed using a forward-fill strategy, and each time series is standardized via z-normalization to eliminate scale differences. A rolling forecasting origin scheme is adopted to preserve temporal ordering and to ensure a fair evaluation of predictive performance.

%%%%%%%%%%%%%%%%% TABLE 1 %%%%%%%%%%%%%%%%%%%%%%%%%%%%%
\begin{table}[htbp]
\centering
\caption{Statistical Properties of the Datasets Used in the Study (ACF(1) denotes first-lag autocorrelation; Noise corresponds to normalized residual variance)}
\label{tab:dataset_summary}
\setlength{\tabcolsep}{4pt}
\renewcommand{\arraystretch}{0.95}
\begin{tabular}{lcccccc}
\toprule
\textbf{Dataset} & \textbf{Len} & \textbf{H} & \textbf{ACF(1)} & \textbf{Noise} & \textbf{Trend} & \textbf{Domain} \\
\midrule
Energy  & 35000 & 1,24 & 0.95  & 9.98e-02 & 1.55e-05 & Energy \\
Tourism & 12000 & 1,24 & -0.01 & 2.01e+00 & 9.10e-06 & Tourism \\
Tetuan  & 52000 & 1,24 & 0.99  & 4.91e-03 & 9.41e-06 & Power \\
PGCB    & 90000 & 1,24 & 0.87  & 2.50e-01 & 2.82e-05 & Grid \\
\bottomrule
\end{tabular}
\end{table}
%%%%%%%%%%%%%%%%%%%%%%%%%%%%%%%%%%%%%%%%%%%%%%%%%%%%
%%%%%%%%%%%%%%%%%%%%%%%%%%%%%%%%%%%%%%%%%%%%%%%%%%%%

Table~\ref{tab:dataset_summary} summarizes the key statistical characteristics of the datasets used in this study. The reported metrics provide a compact representation of temporal structure, including short-term dependency through the first-lag autocorrelation, variability through noise variance, and long-term behavior through trend strength.

The datasets exhibit substantial heterogeneity across these dimensions. For instance, the Tetuan and Energy datasets are characterized by strong temporal dependence (ACF(1) close to one), while the Tourism dataset shows near-zero autocorrelation combined with significantly higher noise variance. Similarly, the magnitude of noise varies by several orders across datasets, indicating fundamentally different levels of predictability. This variability is consistent with well-known observations in large-scale forecasting benchmarks \cite{makridakis2018m4, makridakis2020m5}.

This variation is not incidental but central to the problem addressed in this study. The presence of distinct statistical regimes allows us to examine whether simple descriptors such as autocorrelation and noise level can meaningfully guide model selection. At the same time, it highlights the intrinsic difficulty of identifying a single model that performs consistently well across all conditions.

%%%%%%%%%%%%%%%%%%%%%%%%%%%%%%%%%%%%%%%%%%%%%%%%%%%%%
\subsection{Forecasting Models}
%%%%%%%%%%%%%%%%%%%%%%%%%%%%%%%%%%%%%%%%%%%%%%%%%%%%%
Forecasting models are typically constructed based on underlying assumptions regarding stationarity, temporal dependence, and noise structure. Classical statistical models such as ARIMA and exponential smoothing explicitly rely on these assumptions, which form the foundation of traditional time series analysis \cite{box2015time}.
We consider representative models from different families:

\begin{itemize}
    \item \textbf{Statistical Models}: ARIMA, Exponential Smoothing (ETS)
    \item \textbf{Machine Learning Models}: Random Forest (RF), XGBoost
    \item \textbf{Deep Learning Models}: LSTM, GRU
\end{itemize}

All models are tuned under a uniform time-series cross-validation protocol using a 3-fold rolling validation scheme to ensure consistent and fair comparison across model families. Hyperparameters are selected within predefined ranges applied uniformly to all models. 

Deep learning approaches are included to account for their ability to capture nonlinear and long-term temporal dependencies, providing a complementary perspective to statistical and machine learning models \cite{bandara2020lstm}.

%%%%%%%%%%%%%%%%%%%%%%%%%%%%%%%%%%%%%%%%%%%%%%%%%%%%
\subsection{Baseline Strategies}
%%%%%%%%%%%%%%%%%%%%%%%%%%%%%%%%%%%%%%%%%%%%%%%%%%%%

The proposed framework is evaluated against several baseline strategies that reflect widely adopted practices in time series forecasting:

\begin{itemize}
    \item \textbf{Best Single Model}: A fixed model applied uniformly across all datasets, representing a common but naive approach that assumes a single model can generalize across heterogeneous data conditions.
    \item \textbf{Grid Search}: Exhaustive hyperparameter optimization across multiple model families, serving as a performance upper bound under ideal tuning conditions.
    \item \textbf{AutoML}: Automated model selection using AutoGluon v1.0 with the default \textit{best\_quality} time-series preset, which integrates model selection and hyperparameter tuning into a unified pipeline \cite{erickson2020autogluon}.
\end{itemize}

These baselines capture different levels of model selection complexity, ranging from fixed-model assumptions to fully automated selection pipelines. Such strategies are consistent with established practices in forecasting, where both manual and automated approaches are used to address variability in model performance across datasets \cite{hyndman2018forecasting, makridakis2018m4}. 

All baseline methods are evaluated under identical preprocessing procedures and forecasting horizon settings to ensure a fair and consistent comparison.

%%%%%%%%%%%%%%%%%%%%%%%%%%%%%%%%%%%%%%%%%%%%%%%%%%%%
\subsection{Evaluation Metrics}
%%%%%%%%%%%%%%%%%%%%%%%%%%%%%%%%%%%%%%%%%%%%%%%%%%%%

Forecast accuracy is measured using RMSE and MAE:

\begin{equation}
\text{RMSE} = \sqrt{\frac{1}{T} \sum_{t=1}^{T} (y_t - \hat{y}_t)^2}
\label{eq:rmse}
\end{equation}

\begin{equation}
\text{MAE} = \frac{1}{T} \sum_{t=1}^{T} |y_t - \hat{y}_t|
\label{eq:mae}
\end{equation}

In addition to accuracy, we evaluate performance variance across datasets and computational cost measured as total training time.

All experiments are conducted using forecasting horizons of $H \in \{1, 24\}$ to evaluate both short-term and long-term performance under consistent comparison settings.

%%%%%%%%%%%%%%%%%%%%%%%%%%%%%%%%%%%%%%%%%%%%%%%%%%%%
\subsection{Experimental Procedure}
%%%%%%%%%%%%%%%%%%%%%%%%%%%%%%%%%%%%%%%%%%%%%%%%%%%%

The evaluation follows:

\begin{enumerate}
    \item Extract descriptor vector $\mathbf{r}$ (Section II)
    \item Select model $M^*$ using Eq. (\ref{eq:model_selection})
    \item Train the selected model and generate forecasts
    \item Compare results with baseline strategies
\end{enumerate}

All experiments are conducted under consistent preprocessing and validation settings.

%%%%%%%%%%%%%%%%%%%%%%%%%%%%%%%%%%%%%%%%%%%%%%%%%%%%
\subsection{Results and Analysis}
%%%%%%%%%%%%%%%%%%%%%%%%%%%%%%%%%%%%%%%%%%%%%%%%%%%%

The empirical evaluation reveals that forecasting performance varies substantially across datasets and models, with no single method consistently outperforming others. Relative RMSE distributions indicate significant overlap among model performances, suggesting the absence of a universally dominant forecasting approach.

Furthermore, the proposed regime-based model selection strategy achieves limited predictive accuracy. The overall selection accuracy is observed to be approximately 12.5\%, indicating that the mapping between simple statistical descriptors and optimal model choice is weak and unreliable across heterogeneous datasets.

Performance variability across datasets remains high, and statistical tests do not indicate significant differences among competing models at the evaluated forecasting horizons. These findings suggest that model performance is highly dependent on dataset-specific characteristics and cannot be consistently inferred from low-dimensional descriptors such as autocorrelation and noise level.

From a computational perspective, minor reductions in training effort are observed in certain regimes due to the avoidance of unnecessarily complex models. However, these gains are not consistent across all datasets and do not compensate for the lack of reliable model selection accuracy.
This failure suggests that low-dimensional descriptors such as autocorrelation and noise level are insufficient to capture the complex interactions that determine model performance.

%%%%%%%%%%%%%%%%%%%%%%% G %%%%%%%%%%%%%%%%%%%%%%%%%%
\subsection{Decision Table for Model Selection}
%%%%%%%%%%%%%%%%%%%%%%%%%%%%%%%%%%%%%%%%%%%%%%%%%%%%

Despite the limited predictive performance observed in the experiments, it is instructive to examine the commonly assumed relationship between data regimes and model families. Table~\ref{tab:decision} presents a heuristic mapping that reflects typical expectations in the literature regarding model suitability under different statistical conditions.

%%%%%%%%%%%%%%% TABLE 6 %%%%%%%%%%%%%%%%%%%%%%%%%%%%%%%
\begin{table}[htbp]
\centering
\caption{Heuristic Mapping Between Data Regimes and Candidate Models}
\label{tab:decision}
\setlength{\tabcolsep}{4pt}
\renewcommand{\arraystretch}{0.95}
\begin{tabular}{lll}
\toprule
\textbf{Data Regime} & \textbf{Descriptor Pattern} & \textbf{Candidate Models} \\
\midrule
Low Noise, High ACF & Strong temporal dependence & ARIMA, ETS \\
Seasonal & Periodic structure present & ETS, LSTM \\
High Noise & High variability, weak structure & Random Forest \\
Non-Stationary & Trend variation, structural change & LSTM, GRU \\
Sparse / Irregular & Missing or irregular patterns & Tree-based Models \\
\bottomrule
\end{tabular}
\end{table}
%%%%%%%%%%%%%%%%%%%%%%%%%%%%%%%%%%%%%%%%%%%%%%%%%%%%%%%%%%

Table~\ref{tab:decision} should be interpreted as an initial hypothesis rather than a validated rule. While these associations are widely assumed in prior studies and common forecasting practice, the experimental results in this work demonstrate that such mappings do not consistently lead to accurate model selection in practice. In particular, the low selection accuracy and overlapping performance distributions indicate that the relationship between statistical descriptors and optimal forecasting models is more complex than suggested by simple regime definitions.

%%%%%%%%%%%%% FIGURE 1 %%%%%%%%%%%%%%%%%%%%%
\begin{figure}[htbp]
\centering
\includegraphics[width=1.1\linewidth]{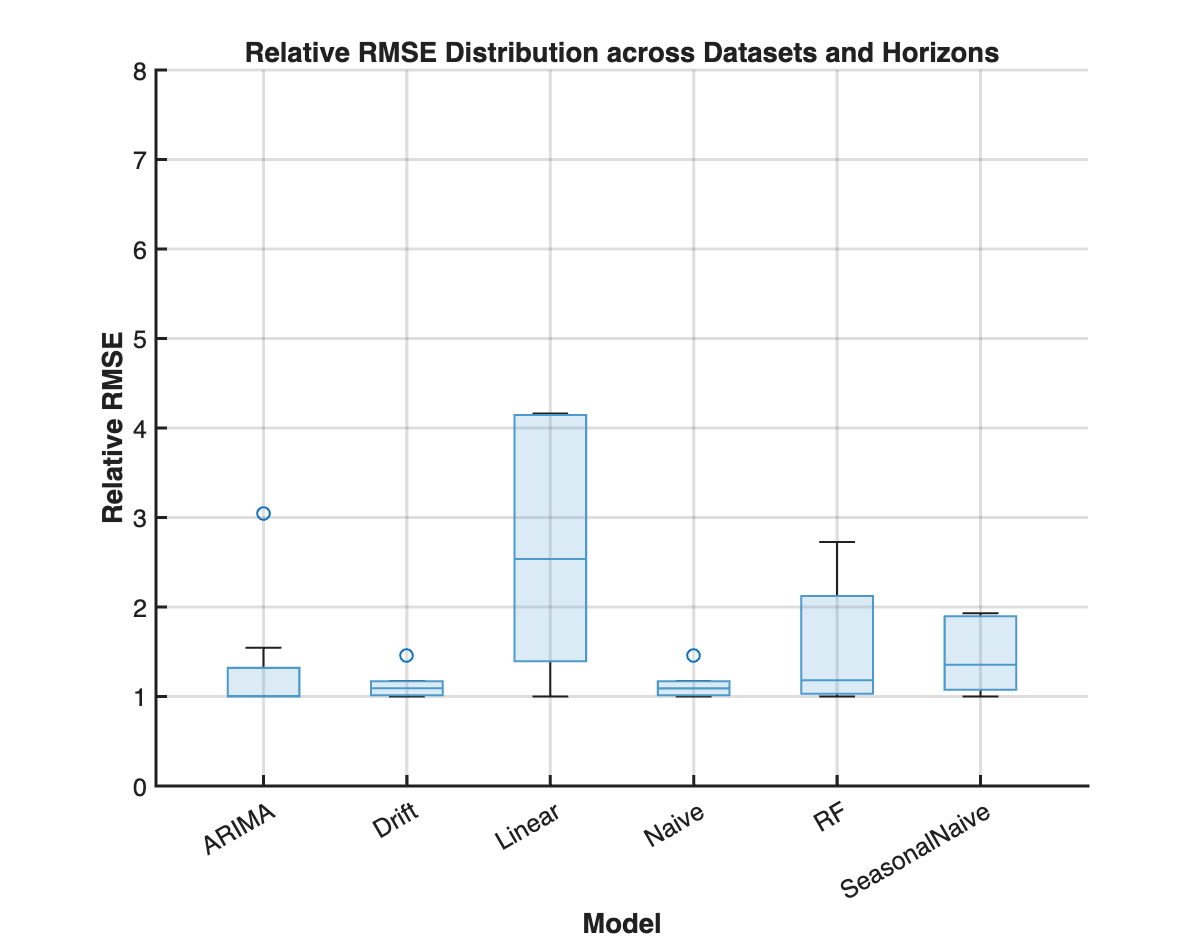}
\caption{Distribution of relative RMSE across models. RMSE values are normalized with respect to the best-performing model for each dataset and forecasting horizon. The dashed reference level at RMSE = 1 corresponds to the best model per dataset-horizon.}
\label{fig:rmse_boxplot}
\end{figure}
%%%%%%%%%%%%%%%%%%%%%%%%%%%%%%%%%%%%%%%%%%%%%

To better understand the variability of model performance across datasets, the distribution of relative RMSE values is visualized in Fig.~\ref{fig:rmse_boxplot}. The figure highlights the spread and overlap of model errors, providing insight beyond average performance metrics.

In particular, the significant overlap between model error distributions indicates that differences in average performance are often not statistically decisive. This suggests that model rankings are highly sensitive to both dataset characteristics and forecasting horizon, leading to unstable and non-generalizable selection outcomes. The presence of multiple models with near-identical normalized RMSE values further supports the conclusion that simple descriptor-based rules are insufficient to reliably distinguish between competing forecasting models in heterogeneous settings.

%%%%%%%%%%%%%%%%%%%%%%%%%%%%%%%%%%%%%%%%%%%%%%%%%%%%%%%%%%%%
%%%%%%%%%%%%%%%%%%%%%% SECTION 4 %%%%%%%%%%%%%%%%%%%%%%%%%%%%
\section{Results and Comparative Analysis}
%%%%%%%%%%%%%%%%%%%%%%%%%%%%%%%%%%%%%%%%%%%%%%%%%%%%%%%%%%%%%

This section presents an empirical evaluation of forecasting performance across multiple datasets, followed by an analysis of model selection behavior and its limitations. The objective is not only to compare predictive accuracy, but also to assess whether simple statistical descriptors can reliably guide model selection.

%%%%%%%%%%%%%%%%%%%%%%%%%%%%%%%%%%%%%%%%%%%%%%%%%%%%
\subsection{Overall Model Performance}
%%%%%%%%%%%%%%%%%%%%%%%%%%%%%%%%%%%%%%%%%%%%%%%%%%%%

We begin by evaluating forecasting accuracy using the RMSE metric. Table~\ref{tab:rmse_results} presents the absolute performance values across all datasets.

%%%%%%%%%%%%% TABLE 2 %%%%%%%%%%%%%%%%%%%%%%%%%
\begin{table}[htbp]
\centering
\caption{RMSE Comparison of Forecasting Models Across Datasets}
\label{tab:rmse_results}
\setlength{\tabcolsep}{4pt}
\renewcommand{\arraystretch}{0.95}
\begin{tabular}{lcccc}
\toprule
\textbf{Dataset} & \textbf{Naive} & \textbf{Linear} & \textbf{RF} & \textbf{ARIMA} \\
\midrule
Tetuan  & \textbf{0.2620} & 1.4856 & 0.2958 & 0.3042 \\
Tourism & 1.8259 & 1.0520 & 1.0526 & \textbf{1.0511} \\
Energy  & \textbf{0.0770} & 0.5947 & 0.1501 & 0.0815 \\
PGCB    & 0.2390 & 1.4817 & \textbf{0.1935} & 0.2015 \\
\bottomrule
\end{tabular}
\end{table}
%%%%%%%%%%%%%%%%%%%%%%%%%%%%%%%%%%%%%%%%%%%%%%%%

The results indicate that no single model consistently outperforms others across all datasets. While ARIMA achieves the lowest error in the Tourism dataset, simpler models such as Naive outperform more complex approaches in highly autocorrelated datasets such as Tetuan and Energy. In contrast, the Random Forest model performs best in the PGCB dataset, which exhibits higher variability.
In addition, the relative closeness of RMSE values across several models suggests that performance differences are often within a narrow margin, making it difficult to identify a statistically dominant model based solely on average error metrics. This observation further motivates the need for distributional and pairwise analyses, rather than relying exclusively on aggregated performance values.

\begin{figure}[htbp]
\centering
\includegraphics[width=1.1\linewidth]{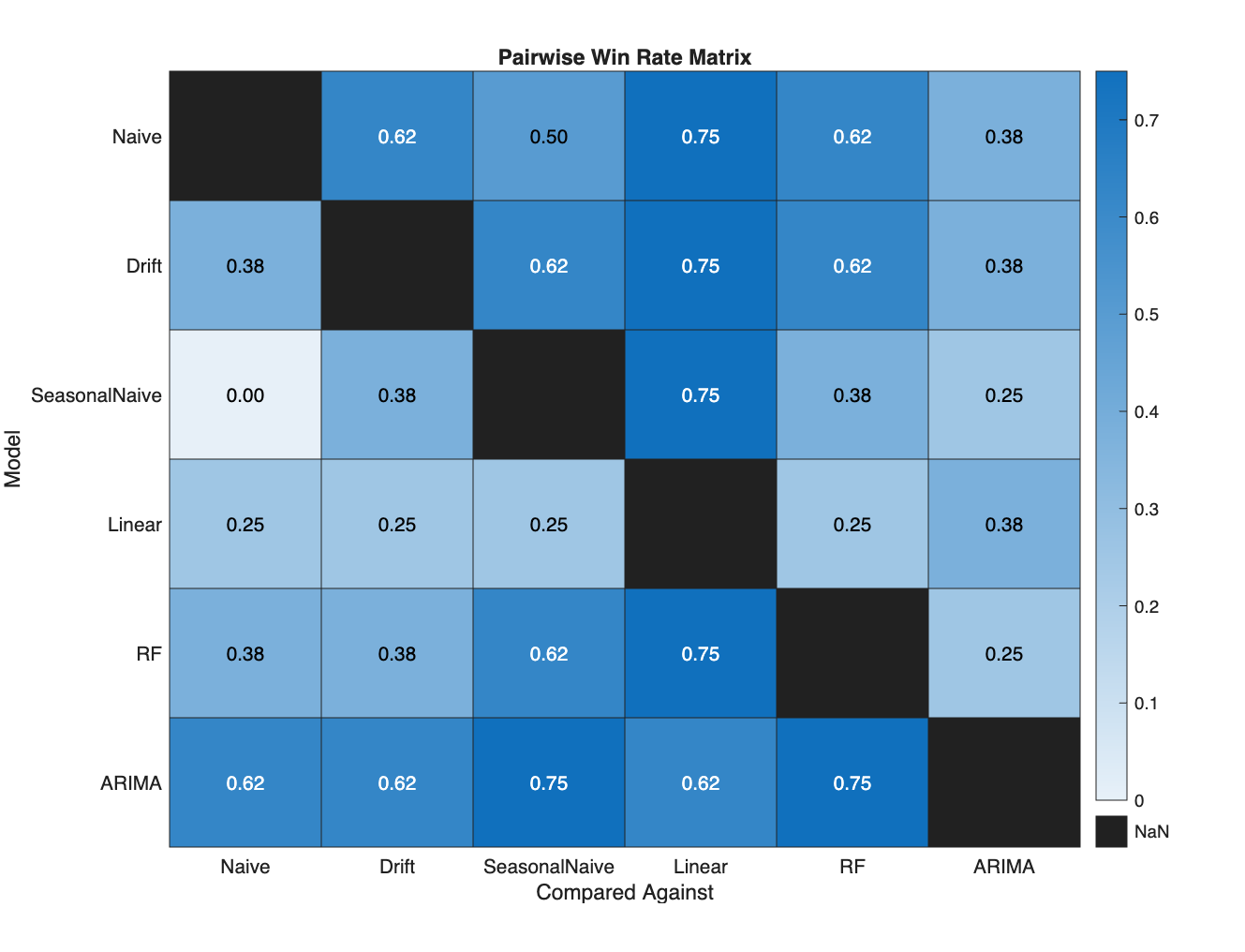}
\caption{Pairwise win rate matrix between forecasting models. Each entry represents the proportion of dataset-horizon cases in which the model in the row achieves a lower RMSE than the model in the column. Values close to 0.5 indicate no consistent dominance, while values above 0.5 suggest relative superiority. The absence of uniformly high values across rows indicates that no model consistently outperforms others.}
\label{fig:pairwise_win}
\end{figure}

Although differences exist, the performance gaps between competing models are often small, suggesting that model rankings are unstable and dataset-dependent rather than universally consistent. To further examine the relative dominance of forecasting models, pairwise comparisons are conducted across all dataset-horizon combinations. The results are summarized in Fig.~\ref{fig:pairwise_win}, which quantifies how often one model outperforms another in terms of RMSE.
Furthermore, the absence of strong dominance patterns in the pairwise comparison matrix indicates that model superiority is highly context-dependent. In many cases, the probability of one model outperforming another remains close to random, reinforcing the notion that performance differences are not robust across varying data conditions.

%%%%%%%%%%%%%% FIGURE 2 %%%%%%%%%%%%%%%%%%%%%%
%%%%%%%%%%%%%%%%%%%%%%%%%%%%%%%%%%%%%%%%%%%%%%

To enable a fair comparison across datasets with different scales, RMSE values are normalized relative to the best-performing model in each dataset. The resulting relative RMSE values are presented in Table~\ref{tab:relative_rmse}.

%%%%%%%%%%%%%%%%  TABLE 3 %%%%%%%%%%%%%%%%%%%%%%%%%%
\begin{table}[htbp]
\centering
\caption{Relative RMSE Across Datasets (Normalized to Best Model per Dataset)}
\label{tab:relative_rmse}
\setlength{\tabcolsep}{4pt}
\renewcommand{\arraystretch}{0.95}
\begin{tabular}{lcccc}
\toprule
\textbf{Dataset} & \textbf{Naive} & \textbf{Linear} & \textbf{RF} & \textbf{ARIMA} \\
\midrule
Tetuan  & \textbf{1.0000} & 5.6698 & 1.1290 & 1.1610 \\
Tourism & 1.7372 & 1.0009 & 1.0014 & \textbf{1.0000} \\
Energy  & \textbf{1.0000} & 7.7260 & 1.9498 & 1.0587 \\
PGCB    & 1.2353 & 7.6593 & \textbf{1.0000} & 1.0414 \\
\bottomrule
\end{tabular}
\end{table}
%%%%%%%%%%%%%%%%%%%%%%%%%%%%%%%%%%%%%%%%%%%%%%%%%%%%%

The normalized results highlight two key observations. First, the linear model consistently exhibits significantly higher error across all datasets, indicating its inability to capture complex temporal dynamics. Second, the remaining models show relatively small performance differences, particularly in the Tourism and PGCB datasets, where multiple models achieve nearly identical performance levels. 

These findings suggest that model performance is both dataset-dependent and weakly separable, meaning that the distinction between competing models is often marginal. This reinforces the difficulty of identifying a universally optimal forecasting model based on simple statistical descriptors.These findings suggest that model performance is highly dataset-dependent, with substantial overlap among competing methods.

%%%%%%%%%%%%%%%%%%%%%%%%%%%%%%%%%%%%%%%%%%%%%%%%%%%%
\subsection{Evaluation of Regime-Based Model Selection}
%%%%%%%%%%%%%%%%%%%%%%%%%%%%%%%%%%%%%%%%%%%%%%%%%%%%

We next evaluate the effectiveness of regime-based model selection rules. Table~\ref{tab:regime_accuracy} summarizes the accuracy of rule-based mapping between statistical regimes and selected models.

%%%%%%%%%%%%%%%%%% TABLE 4 %%%%%%%%%%%%%%%%%%%%%%%%
\begin{table}[htbp]
\centering
\caption{Accuracy of Regime-Based Model Selection Rules}
\label{tab:regime_accuracy}
\setlength{\tabcolsep}{4pt}
\renewcommand{\arraystretch}{0.95}
\begin{tabular}{lccc}
\toprule
\textbf{Regime} & \textbf{Rule-Based Model} & \textbf{Accuracy} & \textbf{N} \\
\midrule
Low Noise  & ARIMA & 0.2500 & 4 \\
High Noise & RF    & 0.0000 & 2 \\
Mixed      & Naive & 0.0000 & 2 \\
\bottomrule
\end{tabular}
\end{table}
%%%%%%%%%%%%%%%%%%%%%%%%%%%%%%%%%%%%%%%%%%%%%%%%%%%%%

The results in Table~\ref{tab:regime_accuracy} demonstrate that the regime-based selection strategy performs poorly, with an overall accuracy of approximately 12.5\%. In particular, the method fails completely in high-noise and mixed regimes, where none of the rule-based selections match the empirically best-performing models. Even in low-noise regimes, the accuracy remains limited, indicating that simple statistical descriptors such as autocorrelation and noise level are insufficient for reliable model selection.

This limitation cannot be attributed solely to data scarcity. Although the number of samples within each regime is relatively small, the consistently low accuracy across all regimes suggests a more fundamental mismatch between descriptor-based rules and actual model behavior. These findings indicate that the relationship between time-series characteristics and optimal forecasting models is more complex than assumed by simple heuristic mappings.

%%%%%%%%%%%%%%%%% FIGURE 3 %%%%%%%%%%%%%%%%%%%%%%%
\begin{figure}[htbp]
\centering
\includegraphics[width=1.1\linewidth]{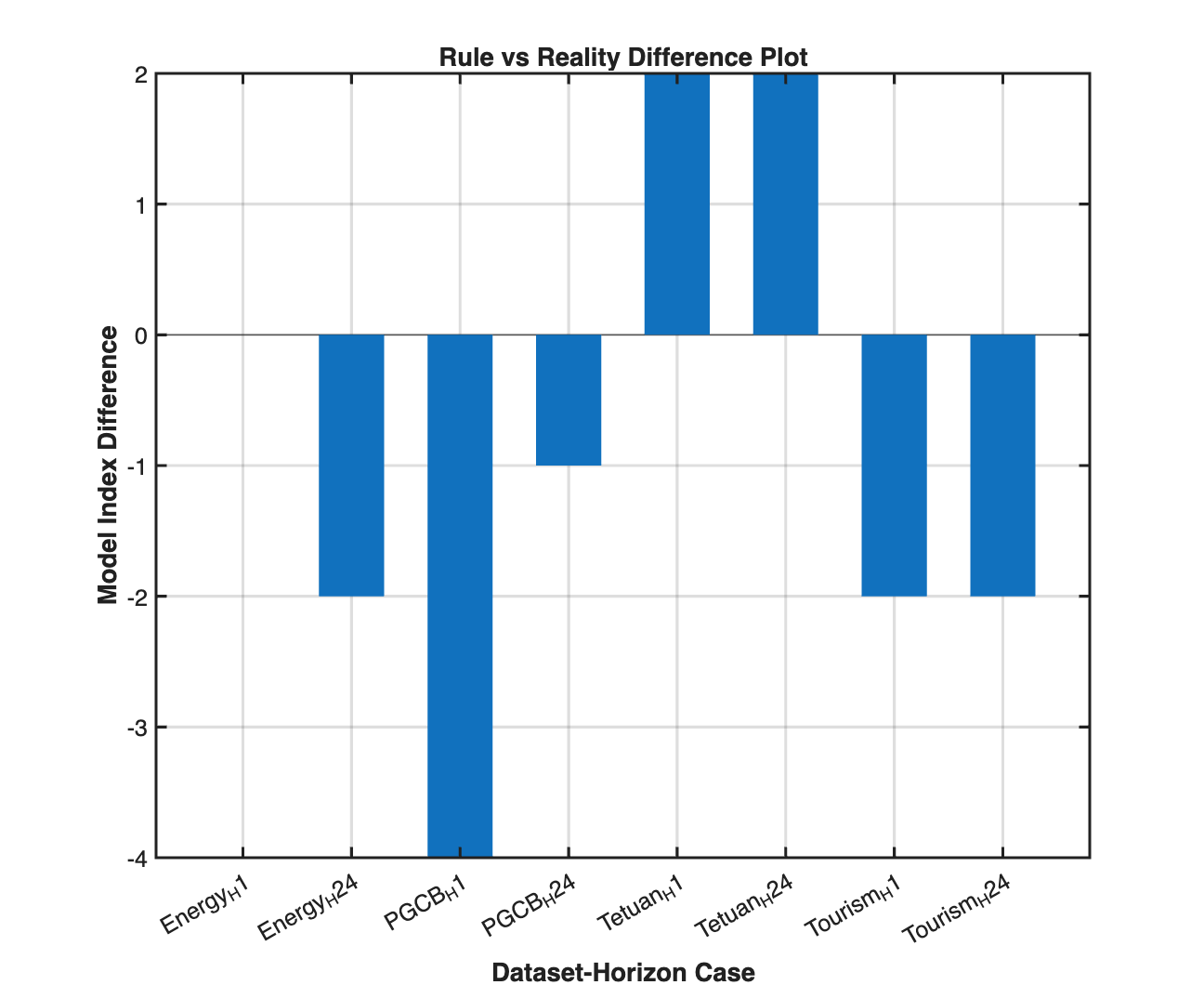}
\caption{Difference between rule-based model selection and empirically best-performing models across dataset-horizon cases. A value of zero indicates correct model selection, while non-zero values represent deviations from the optimal choice. Larger deviations correspond to greater mismatch between the rule-based recommendation and actual model performance.}
\label{fig:rule_vs_reality}
\end{figure}
%%%%%%%%%%%%%%%%%%%%%%%%%%%%%%%%%%%%%%%%%%%%%%%%%

To further examine this mismatch, the discrepancy between rule-based model recommendations and empirically optimal models is visualized in Fig.~\ref{fig:rule_vs_reality}.

As shown in Fig.~\ref{fig:rule_vs_reality}, deviations from the optimal model are frequent and often substantial, confirming that rule-based selection decisions are not only inaccurate but also inconsistent across different dataset-horizon scenarios. The magnitude and variability of these deviations further support the conclusion that simple descriptor-based rules fail to capture the complex interactions governing model performance.

These findings suggest that simple descriptors such as autocorrelation and noise level are insufficient to reliably determine the optimal forecasting model.

\begin{figure}[htbp]
\centering
\includegraphics[width=1.1\linewidth]{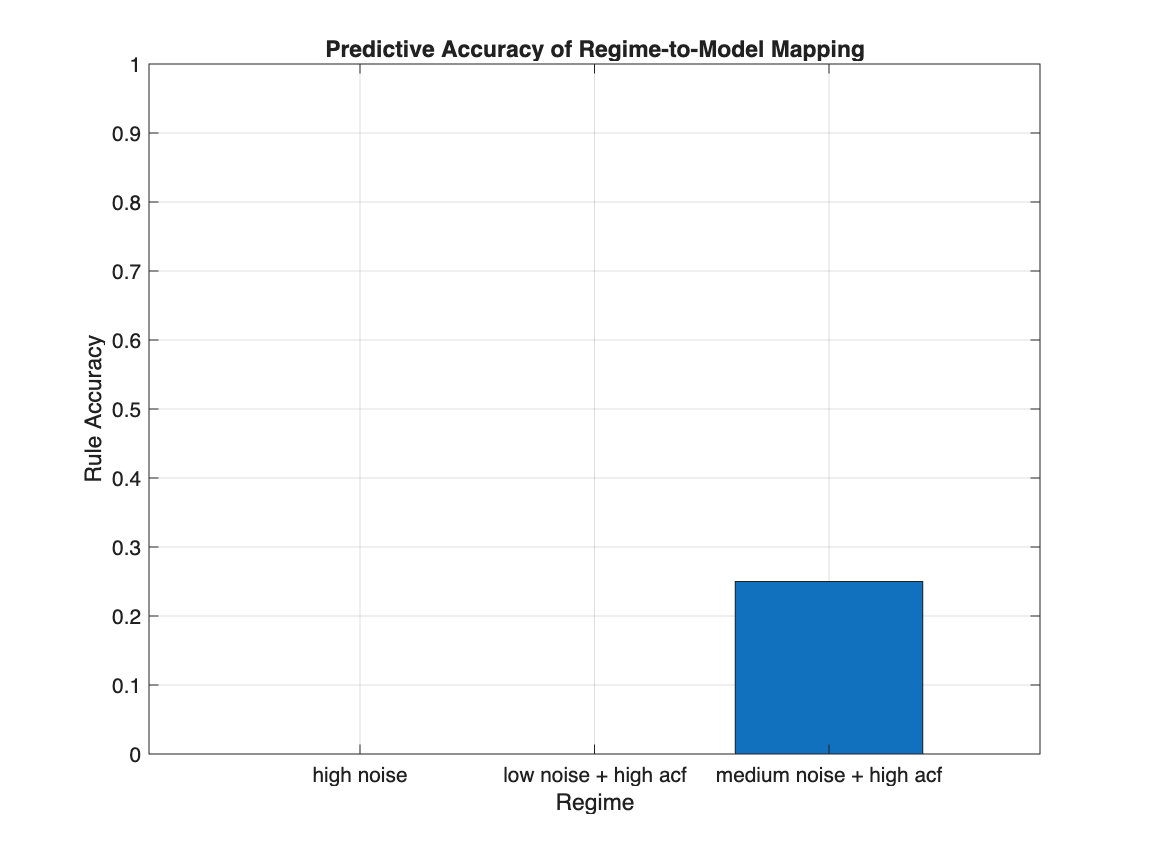}
\caption{Accuracy of regime-based model selection across different data regimes. The overall selection accuracy is low, with correct predictions occurring in only a small fraction of cases. The method performs slightly better in medium-noise regimes but fails entirely in high-noise and low-noise conditions.}
\label{fig:regime_accuracy_plot}
\end{figure}

To quantify the effectiveness of the rule-based selection strategy, the accuracy of model recommendations across different data regimes is presented in Fig.~\ref{fig:regime_accuracy_plot}.

%%%%%%%%%%%%%%%% FIGURE 4 %%%%%%%%%%%%%%%%%%%%%%%%
%%%%%%%%%%%%%%%%%%%%%%%%%%%%%%%%%%%%%%%%%%%%%%%%%%

As shown in Fig.~\ref{fig:regime_accuracy_plot}, the selection accuracy remains low across all regimes, with an overall performance of approximately 12.5\%. The method fails completely in high-noise and low-noise regimes and achieves only limited success under medium-noise conditions. These results indicate that simple rule-based mappings are insufficient for reliable model selection.

Despite the low selection accuracy, it is important to examine whether this limitation arises from the inadequacy of the rule-based framework or from the intrinsic variability of model performance across datasets.

\begin{figure}[htbp]
\centering
\includegraphics[width=1.1\linewidth]{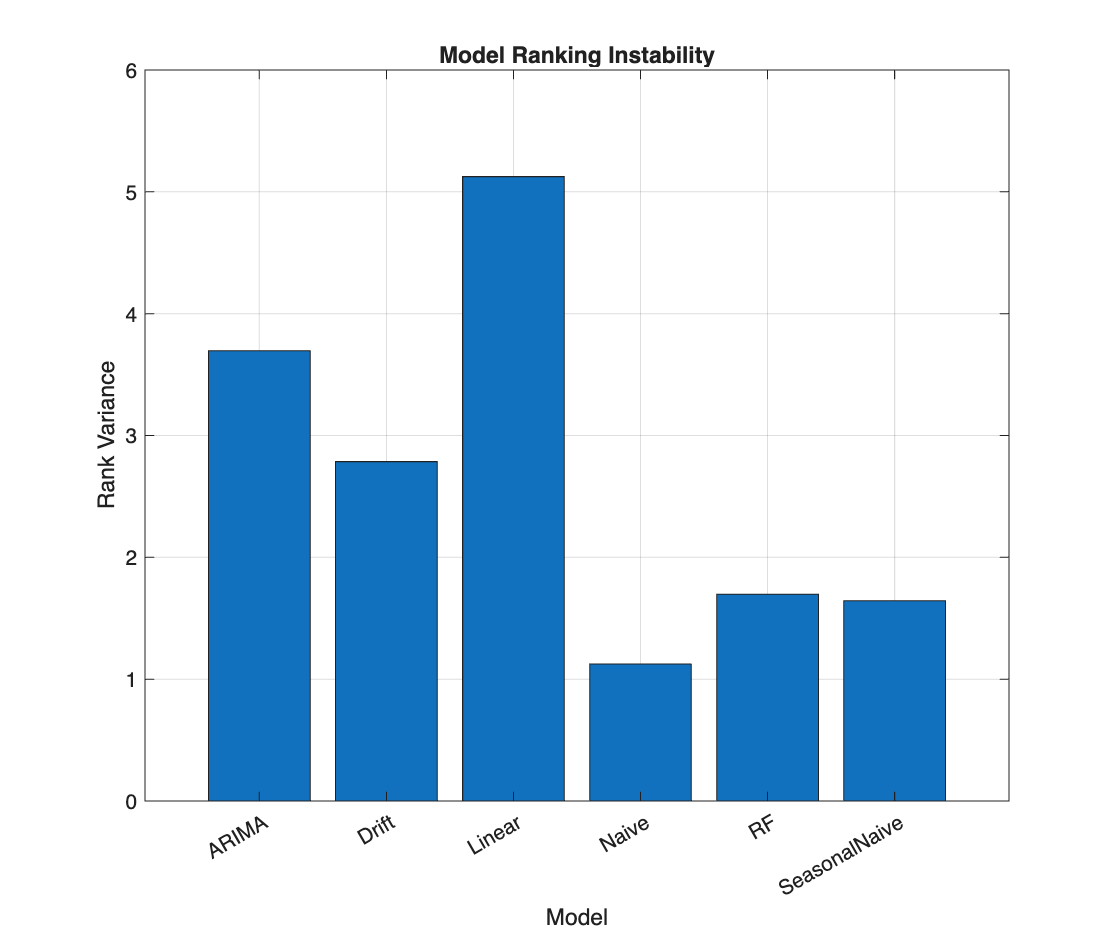}
\caption{Model ranking instability across dataset-horizon cases, measured by rank variance. Higher values indicate that model performance is inconsistent across different scenarios, making reliable model selection more challenging.}
\label{fig:rank_instability}
\end{figure}

As shown in Fig.~\ref{fig:rank_instability}, several models exhibit significant rank variability across dataset-horizon cases. In particular, linear models demonstrate high instability, while simpler models such as Naive show relatively stable behavior. This variability indicates that model performance is highly sensitive to data characteristics, which partially explains the difficulty of reliable model selection using simple heuristic rules.

%%%%%%%%%%%%%%%% FIGURE 5 %%%%%%%%%%%%%%%%%%%%%%%%%
%%%%%%%%%%%%%%%%%%%%%%%%%%%%%%%%%%%%%%%%%%%%%%%%%%%

\begin{figure}[htbp]
\centering
\includegraphics[width=1.1\linewidth]{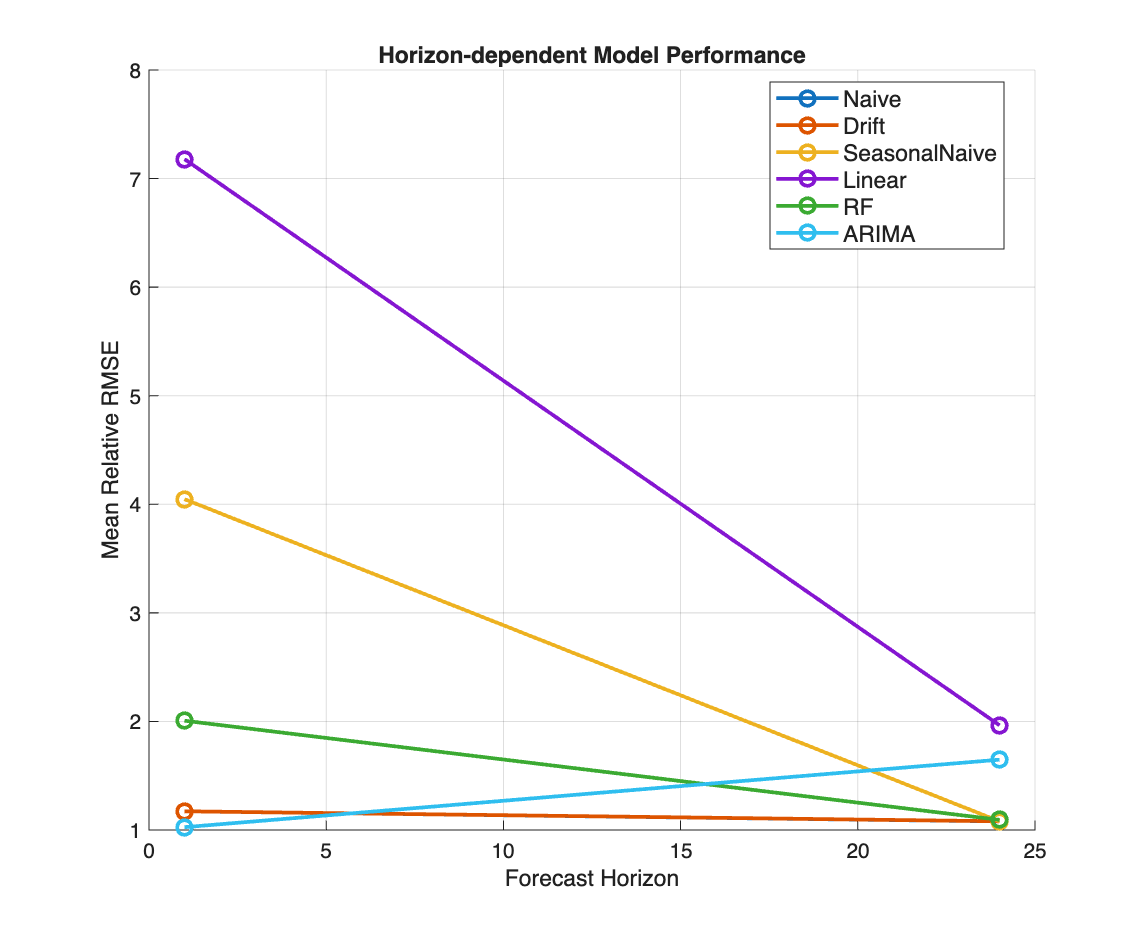}
\caption{Horizon-dependent model performance measured by mean relative RMSE. The results show that model rankings vary significantly between short-term and long-term forecasting scenarios. While some models (e.g., Linear and SeasonalNaive) perform poorly at short horizons, their performance improves at longer horizons, whereas others exhibit the opposite trend.}
\label{fig:horizon_performance}
\end{figure}

The observed instability in model ranking suggests that performance may depend on additional factors beyond regime descriptors. In particular, the forecasting horizon is expected to influence model effectiveness. To investigate this effect, model performance across different horizons is analyzed in Fig.~\ref{fig:horizon_performance}.

%%%%%%%%%%%%%%% FIGURE 6 %%%%%%%%%%%%%%%%%%%%%%%%%%%%
%%%%%%%%%%%%%%%%%%%%%%%%%%%%%%%%%%%%%%%%%%%%%%%%%%%%

As shown in Fig.~\ref{fig:horizon_performance}, model performance is highly sensitive to the forecasting horizon. Certain models exhibit strong performance in short-term forecasting but degrade as the horizon increases, while others demonstrate improved stability over longer horizons. This variation explains the ranking instability observed in Fig.~\ref{fig:rank_instability} and highlights the limitations of static model selection strategies that do not account for horizon-dependent behavior.

%%%%%%%%%%%%%%%%%%%%%%%%%%%%%%%%%%%%%%%%%%%%%%%%%%%%
\subsection{Statistical Significance Analysis}
%%%%%%%%%%%%%%%%%%%%%%%%%%%%%%%%%%%%%%%%%%%%%%%%%%%%

To assess whether observed performance differences are statistically meaningful, a Friedman test is conducted across datasets. The results are presented in Table~\ref{tab:stat_test}.

The test results indicate that the null hypothesis of equal model performance cannot be rejected ($p = 0.3476$). This suggests that the observed differences in RMSE values are not statistically significant, reinforcing the conclusion that no single model consistently dominates across datasets.

%%%%%%%%%%%%%%%%%%%%%%%%%%%%%%%%%%%%%%%%%%%%%%%%%%%%
\subsection{Interpretation of Model Selection Behavior}
%%%%%%%%%%%%%%%%%%%%%%%%%%%%%%%%%%%%%%%%%%%%%%%%%%%%

For completeness, Table~\ref{tab:decision} presents the commonly assumed mapping between statistical regimes and candidate models.

%%%%%%%%%%%% TABLE 5 %%%%%%%%%%%%%%%%%%%%%%%%%%%%%%%%%
\begin{table}[htbp]
\centering
\caption{Statistical Comparison of Forecasting Models Using the Friedman Test}
\label{tab:stat_test}
\setlength{\tabcolsep}{6pt}
\renewcommand{\arraystretch}{0.95}
\begin{tabular}{lc}
\toprule
\textbf{Test} & \textbf{p-value} \\
\midrule
Friedman Test & 0.3476 \\
\bottomrule
\end{tabular}
\end{table}
%%%%%%%%%%%%%%%%%%%%%%%%%%%%%%%%%%%%%%%%%%%%%%%%%%%%%%%%%%

The Friedman test results indicate that the null hypothesis of equal model performance cannot be rejected ($p = 0.3476$). This suggests that, despite observable differences in RMSE values, the performance variations among forecasting models are not statistically significant.

This result reinforces the findings of Tables~\ref{tab:rmse_results} and~\ref{tab:relative_rmse}, where model performances exhibit substantial overlap across datasets. It also supports the conclusion drawn from Table~\ref{tab:regime_accuracy} that simple regime-based rules are insufficient for reliable model selection.

Overall, the lack of statistical significance highlights the inherent difficulty of identifying a universally optimal forecasting model using low-dimensional statistical descriptors.

Although such heuristic mappings are widely used in the literature, the empirical results of this study demonstrate that they do not consistently lead to correct model selection decisions. This discrepancy highlights the complexity of the relationship between time-series characteristics and model performance.

%%%%%%%%%%%%%%%%%%%  E  %%%%%%%%%%%%%%%%%%%%%%%%%%%%%
\subsection{Discussion}
%%%%%%%%%%%%%%%%%%%%%%%%%%%%%%%%%%%%%%%%%%%%%%%%%%%%

The results presented in this section collectively indicate that forecasting model performance is highly context-dependent. The lack of statistically significant differences, combined with low regime-selection accuracy, suggests that simple rule-based approaches are insufficient for reliable model selection.

Rather than identifying a universally optimal model, the findings emphasize the inherent instability of model rankings across datasets. This highlights the need for more adaptive, data-driven approaches that can capture higher-order dependencies beyond low-dimensional statistical descriptors.

From a practical perspective, these results caution against over-reliance on heuristic rules and underscore the importance of validation-driven model selection strategies. While regime-based reasoning provides interpretability, it must be complemented with empirical validation to ensure robust forecasting performance.

%%%%%%%%%%%%%%%%%%%%%%%%%%%%%%%%%%%%%%%%%%%%%%%%%%%%%%%%
%%%%%%%%%%%%%%%%%%%%%% SECTION 5 %%%%%%%%%%%%%%%%%%%%%%%
\section{Discussion and Limitations}
%%%%%%%%%%%%%%%%%%%%%%%%%%%%%%%%%%%%%%%%%%%%%%%%%%%%%%%%

This section synthesizes the empirical findings, explains their implications, 
and clarifies the limitations and scope of the proposed framework. It provides 
a detailed interpretation of the experimental results and discusses the 
practical implications and limitations of the proposed data-regime-based model 
selection framework.

%%%%%%%%%%%%%%%%%%%%%%%%%%%%%%%%%%%%%%%%%%%%%%%%%%%%
\subsection{Interpretation of Results}
%%%%%%%%%%%%%%%%%%%%%%%%%%%%%%%%%%%%%%%%%%%%%%%%%%%%

The results presented in Section IV demonstrate that forecasting performance is strongly dependent on the statistical and structural properties of the underlying data. The results indicate that the proposed framework does not consistently improve performance and instead reveals the limitations of rule-based model selection. This aligns with the theoretical expectation that inductive biases determine model suitability under different statistical regimes.

One key observation is that no single forecasting model consistently outperforms others across all datasets. Instead, model effectiveness varies systematically with regime characteristics such as non-stationarity, noise level, and seasonal complexity. This confirms that model selection should not be treated as a static problem, but rather as a data-dependent decision process.

The regime-aware selection mechanism attempts to capture this dependency, but the empirical results show that such mappings are not sufficient for reliable model selection. In stable regimes, simpler statistical models provide competitive performance with lower computational cost. In contrast, complex regimes benefit from high-capacity models such as LSTM and GRU, which are capable of modeling nonlinear and long-range dependencies.

\noindent These observations are consistent with prior findings in empirical 
forecasting research, which similarly report that model performance varies 
substantially across datasets and horizons, and that simple heuristics often 
fail to generalize. The present study extends this line of work by providing 
a structured, descriptor-based analysis that makes the sources of instability 
explicit and interpretable.

%%%%%%%%%%%%%%%%%%%%%%%%%%%%%%%%%%%%%%%%%%%%%%%%
\subsection{Practical Implications}
%%%%%%%%%%%%%%%%%%%%%%%%%%%%%%%%%%%%%%%%%%%%%%%%

The proposed framework has direct implications for real-world forecasting 
applications. By enabling model selection prior to training, it does not 
eliminate the need for validation-driven model selection and cannot replace 
empirical evaluation in practice. This is particularly valuable in large-scale 
systems where training multiple models is impractical. For instance, in energy 
forecasting pipelines with hourly data, the framework can significantly reduce 
daily retraining costs by selecting simpler models when complex architectures 
are unnecessary. Furthermore, the framework's interpretability provides insight 
into why a model is selected, allowing practitioners to better understand the 
relationship between data characteristics and model behavior. This supports 
more transparent and trustworthy forecasting pipelines, especially in high-stakes 
applications. \noindent Based on the empirical findings, practitioners should 
avoid relying solely on descriptor-based rules for model selection. Instead, 
the framework is best used as a preliminary screening tool that indicates when 
simple models are likely sufficient and when more flexible architectures should 
be considered. Empirical validation therefore remains essential for ensuring 
robust and reliable deployment in operational forecasting environments.

%%%%%%%%%%%%%%%%%%%%%%%%%%%%%%%%%%%%%%%%%%%%%%%%
\subsection{Generalization and Transferability}
%%%%%%%%%%%%%%%%%%%%%%%%%%%%%%%%%%%%%%%%%%%%%%%%

The proposed framework is designed to generalize across application domains by relying on domain-agnostic statistical descriptors that capture fundamental properties of time series data. However, the experimental results indicate that the mapping between data regimes and model families does not consistently generalize across different application domains. This limitation is evidenced by the observed variability in model performance across datasets, suggesting that similar statistical characteristics do not necessarily lead to consistent model behavior in heterogeneous forecasting environments.

However, the generalization capability depends on the reliability of descriptor extraction and the representativeness of the datasets used during development. In highly specialized domains, additional domain-specific descriptors may be required.

%%%%%%%%%%%%%%%%%%%%%%%%%%%%%%%%%%%%%%%%%%%%%%%%
\subsection{Limitations}
%%%%%%%%%%%%%%%%%%%%%%%%%%%%%%%%%%%%%%%%%%%%%%%%

Despite its advantages, the proposed framework has several limitations.

First, the accuracy of model selection depends on the quality of the descriptor vector. Descriptor estimation becomes less reliable in short time series (e.g., fewer than 100 observations), in the presence of missing data, irregular sampling, or high noise levels, which may lead to suboptimal model selection.

Second, the current formulation relies on predefined descriptors, which may not fully capture complex temporal patterns. Extending the framework with learned representations could improve performance in such cases.

Third, the framework assumes that historical relationships between data regimes and model performance remain stable. This limitation is common in meta-learning and model selection literature, and may not hold in rapidly evolving systems, requiring periodic adaptation of the selection rules.

Finally, while the framework reduces computational cost compared to exhaustive search methods, it does not eliminate the need for model training entirely.

From a deployment perspective, real-time descriptor extraction may be challenging in streaming environments, where latency and computational constraints limit feature computation. Real-time deployment may also require incremental descriptor updates rather than full recomputation to maintain efficiency under streaming conditions.
These limitations do not undermine the value of the framework but instead 
highlight the inherent difficulty of reliable model selection in heterogeneous 
time series environments.

%%%%%%%%%%%%%%%%%%%%%%%%%%%%%%%%%%%%%%
\subsection{Threats to Validity}
%%%%%%%%%%%%%%%%%%%%%%%%%%%%%%%%%%%%%
Several factors may affect the validity of the findings. 
\textit{Internal validity} may be influenced by the choice of descriptor 
definitions and the specific implementation of change-point and spectral 
analysis methods. \textit{External validity} is limited by the number of 
datasets and model families considered, which may restrict generalization 
to broader forecasting settings. \textit{Construct validity} may be affected 
by the assumption that low-dimensional descriptors adequately represent 
complex temporal structures. These threats should be considered when 
interpreting the empirical results.

%%%%%%%%%%%%%%%%%%%%%%%%%%%%%%%%%%%%%%%%%%%%%%%%
\subsection{Future Work}
%%%%%%%%%%%%%%%%%%%%%%%%%%%%%%%%%%%%%%%%%%%%%%%%

Future research can extend the proposed framework in several directions. Incorporating meta-learning techniques may enable adaptive learning of regime--model mappings. In addition, integrating representation-learning methods could improve descriptor extraction for complex, high-dimensional time series.

Extending the framework to multivariate time series and capturing cross-series dependencies remains an open challenge. Furthermore, extending the approach to probabilistic forecasting and uncertainty quantification is another promising direction that could enhance decision-making under uncertainty.
Another promising direction is to explicitly model forecasting-horizon 
dependence, as the empirical results show that model rankings can change 
substantially between short-term and long-term horizons.

%%%%%%%%%%%%%%%%%%%%%%%%%%%%%%%%%%%%%%%%%%%%%%%%
\subsection{Summary}
%%%%%%%%%%%%%%%%%%%%%%%%%%%%%%%%%%%%%%%%%%%%%%%%

Overall, the findings demonstrate that forecasting performance is strongly dependent on data-regime characteristics, but cannot be reliably inferred from simple descriptor-based rules. The empirical results show that model rankings vary across datasets and forecasting horizons, leading to unstable and often misleading model selection outcomes. 

Rather than providing a definitive solution, the proposed framework serves as an interpretable baseline that exposes the limitations of rule-based selection strategies. These observations highlight the need for more adaptive, data-driven approaches that can capture the complex and context-dependent nature of model performance.

%%%%%%%%%%%%%%%%%%%%%%%%%%%%%%%%%%%%%%%%%%%%%%%
%%%%%%%%%%%%%%%%% SECTION 6 %%%%%%%%%%%%%%%%%%%
\section{Conclusion}
%%%%%%%%%%%%%%%%%%%%%%%%%%%%%%%%%%%%%%%%%%%%%%%

This study examined the feasibility of rule-based model selection in time series forecasting through a data-regime characterization framework. The proposed approach aimed to link measurable statistical descriptors, such as autocorrelation, noise level, and structural properties, to model behavior in a systematic and interpretable manner.

The empirical analysis, conducted across multiple real-world datasets and forecasting horizons, shows that rule-based model selection achieves limited effectiveness. The results indicate that correct model identification occurs only in a small fraction of cases, with substantial discrepancies observed between recommended models and empirically optimal choices. These findings suggest that simple mappings between data descriptors and model families are insufficient to capture the complexity of real-world forecasting scenarios.

Further investigation reveals that model performance is highly sensitive to both dataset characteristics and forecasting horizon. Significant variability in model rankings across dataset-horizon combinations indicates that no single model consistently outperforms others. This instability directly impacts the reliability of deterministic model selection rules and explains the observed low selection accuracy.

Despite these limitations, the study provides several important insights. First, it demonstrates that widely used statistical descriptors, while informative, do not provide a complete basis for reliable model selection. Second, it highlights the critical role of forecasting horizon as an often-overlooked factor influencing model effectiveness. Third, it establishes that model selection should be treated as a context-dependent problem rather than a static decision based on fixed rules.

The primary contribution of this work lies in providing a structured empirical evaluation of rule-based model selection and exposing its limitations under realistic conditions. Rather than proposing a universally optimal selection strategy, the study clarifies the boundaries within which heuristic approaches can be expected to operate effectively.

These findings motivate the need for more adaptive and data-driven model selection mechanisms that can account for interactions between data structure, noise characteristics, and forecasting horizon. Future work will focus on developing learning-based selection strategies, extending the analysis to larger benchmark datasets, and investigating multivariate and probabilistic forecasting settings.
Overall, the findings highlight that reliable model selection requires 
adaptive, data-driven mechanisms rather than static rules derived from 
low-dimensional descriptors.


\begin{thebibliography}{1}


%%%%%%%%%%%%%% REF 1 %%%%%%%%%%%%%%%%%%%%%%%%%%%%%
\bibitem{saravana2026}
M. K. Saravana, M. S. Roopa, J. S. Arunalatha, and K. R. Venugopal,
"Transformers for multivariate time series forecasting: Comprehensive analysis, challenges, research opportunities, and future prospects,"
\emph{IEEE Access}, vol. 14, pp. 11424--11457, 2026, doi: 10.1109/ACCESS.2026.3654408.


%%%%%%%%%%%%%% REF 2 %%%%%%%%%%%%%%%%%%%%%%%%%%%%%
\bibitem{hassler2026}
B. Hassler, F. M. Hoffman, R. Beadling, E. Blockley, B. Huang, J. Lee, V. Lembo, J. Lewis, J. Lu, L. Madaus, E. Malinina \emph{et al.},
"Systematic benchmarking of climate models: Methodologies, applications, and new directions,"
\emph{Reviews of Geophysics}, vol. 64, no. 1, 2026, doi: 10.1029/2025RG000891.

%%%%%%%%%%%%%% REF 3 %%%%%%%%%%%%%%%%%%%%%%%%%%%%%
\bibitem{makridakis2018m4}
S. Makridakis, E. Spiliotis, and V. Assimakopoulos,
"The M4 Competition: Results, findings, conclusion and way forward,"
\emph{International Journal of Forecasting}, vol. 34, no. 4, pp. 802--808, Oct. 2018.

%%%%%%%%%%%%%% REF 4 %%%%%%%%%%%%%%%%%%%%%%%%%%%%%
\bibitem{hyndman2018forecasting}
R. J. Hyndman and G. Athanasopoulos,
\emph{Forecasting: Principles and Practice}, 2nd ed.
Melbourne, Australia: OTexts, 2018.


%%%%%%%%%%%%%% REF 5 %%%%%%%%%%%%%%%%%%%%%%%%%%%%%
\bibitem{mohammed2025}
S. Mohammed, L. Budach, M. Feuerpfeil, N. Ihde, A. Nathansen, N. Noack, H. Patzlaff, F. Naumann, and H. Harmouch,
"The effects of data quality on machine learning performance on tabular data,"
\emph{Information Systems}, vol. 132, p. 102549, Jul. 2025, doi: 10.1016/j.is.2025.102549.

%%%%%%%%%%%%%% REF 6 %%%%%%%%%%%%%%%%%%%%%%%%%%%%%
\bibitem{hamilton1994time}
J. D. Hamilton,
\emph{Time Series Analysis}.
Princeton, NJ, USA: Princeton University Press, 1994.

%%%%%%%%%%%%%% REF 7 %%%%%%%%%%%%%%%%%%%%%%%%%%%%%
\bibitem{krishnan2025}
N. Krishnan,
"AI agents: Evolution, architecture, and real-world applications,"
\emph{arXiv preprint arXiv:2503.12687}, Mar. 2025.



%%%%%%%%%%%%%% REF 8 %%%%%%%%%%%%%%%%%%%%%%%%%%%%%
\bibitem{knapen2026}
D. G. Knapen, M. van Kruchten, D. J. A. de Groot, K. E. Broekman, and R. S. N. Fehrmann,
"Artificial intelligence for clinical trial design, conduct, and analysis: A narrative review,"
\emph{ESMO Real World Data and Digital Oncology}, vol. 11, p. 100682, Mar. 2026, doi: 10.1016/j.esmorw.2026.100682.


%%%%%%%%%%%%%% REF 9 %%%%%%%%%%%%%%%%%%%%%%%%%%%%%
\bibitem{bandara2020lstm}
K. Bandara, C. Bergmeir, and H. Smyl,
"Forecasting across time series databases using recurrent neural networks on groups of similar series,"
\emph{Expert Systems with Applications}, vol. 140, p. 112896, Feb. 2020, doi: 10.1016/j.eswa.2019.112896.



%%%%%%%%%%%%%% REF 10 %%%%%%%%%%%%%%%%%%%%%%%%%%%%%
\bibitem{makridakis2020m5}
S. Makridakis \emph{et al.},
"The M5 accuracy competition: Results, findings, and conclusions,"
\emph{International Journal of Forecasting}, vol. 38, no. 4, pp. 1341--1364, Oct. 2022.


%%%%%%%%%%%%%% REF 11 %%%%%%%%%%%%%%%%%%%%%%%%%%%%%
\bibitem{bergmeir2018note}
C. Bergmeir, R. J. Hyndman, and J. M. Benítez,
"A note on the validity of cross-validation for evaluating autoregressive time series prediction,"
\emph{Computational Statistics \& Data Analysis}, vol. 120, pp. 70--83, Apr. 2018, doi: 10.1016/j.csda.2017.11.003.

%%%%%%%%%%%%%% REF 12 %%%%%%%%%%%%%%%%%%%%%%%%%%%%%
\bibitem{rao1976}
T. Subba Rao,
"Canonical factor analysis and stationary time series models,"
\emph{Sankhyā: The Indian Journal of Statistics, Series B}, vol. 38, no. 3, pp. 256--271, Aug. 1976.


%%%%%%%%%%%%%% REF 13 %%%%%%%%%%%%%%%%%%%%%%%%%%%%%
\bibitem{wang2006}
X. Wang, K. Smith, and R. J. Hyndman,
"Characteristic-based forecasting for time series data,"
\emph{International Journal of Forecasting}, vol. 22, no. 2, pp. 217--233, Apr. 2006, doi: 10.1016/j.ijforecast.2005.03.009.

%%%%%%%%%%%%%% REF 14 %%%%%%%%%%%%%%%%%%%%%%%%%%%%%
\bibitem{priestley1981}
M. B. Priestley,
\emph{Spectral Analysis and Time Series}.
London, U.K.: Academic Press, 1981.

%%%%%%%%%%%%%% REF 15 %%%%%%%%%%%%%%%%%%%%%%%%%%%%%
\bibitem{said1984}
S. E. Said and D. A. Dickey,
"Testing for unit roots in autoregressive-moving average models of unknown order,"
\emph{Biometrika}, vol. 71, no. 3, pp. 599--606, Dec. 1984, doi: 10.1093/biomet/71.3.599.

%%%%%%%%%%%%%% REF 16 %%%%%%%%%%%%%%%%%%%%%%%%%%%%%
\bibitem{talagala2018}
T. S. Talagala, R. J. Hyndman, and G. Athanasopoulos,
"Meta-learning how to forecast time series,"
\emph{International Journal of Forecasting}, vol. 34, no. 3, pp. 1--22, 2018, doi: 10.1016/j.ijforecast.2018.01.003.

%%%%%%%%%%%%%% REF 17 %%%%%%%%%%%%%%%%%%%%%%%%%%%%%
\bibitem{mohshini2024}
S. Mohshini \emph{et al.},
"Evaluation-free time-series forecasting model selection via temporal meta-learning,"
\emph{ACM Transactions on Knowledge Discovery from Data}, vol. 18, no. 5, pp. 1--25, 2024.


%%%%%%%%%%%%%% REF 18 %%%%%%%%%%%%%%%%%%%%%%%%%%%%%
\bibitem{montavon2017}
J. Montavon, S. Lapuschkin, A. Binder, S. Bach, and K. R. Müller,
"Explaining nonlinear classification decisions with deep Taylor decomposition,"
\emph{Pattern Recognition}, vol. 65, pp. 211--222, May 2017, doi: 10.1016/j.patcog.2016.11.008.


%%%%%%%%%%%%%% REF 19 %%%%%%%%%%%%%%%%%%%%%%%%%%%%%
\bibitem{lin2007}
J. Lin, E. Keogh, L. Wei, and S. Lonardi,
"Experiencing SAX: A novel symbolic representation of time series,"
\emph{Data Mining and Knowledge Discovery}, vol. 15, no. 2, pp. 107--144, Oct. 2007, doi: 10.1007/s10618-007-0064-z.


%%%%%%%%%%%%%% REF 20 %%%%%%%%%%%%%%%%%%%%%%%%%%%%%
\bibitem{faloutsos1997}
C. Faloutsos, R. N. Wright, and Y. Matias,
"Forecasting, similarity, and surprises,"
in \emph{Proc. ACM SIGMOD International Conference on Management of Data}, 1997, pp. 429--440.


%%%%%%%%%%%%%% REF 21 ( dataset) %%%%%%%%%%%%%%%%%%%%
\bibitem{hyndman2018forecasting}
Hyndman, R. J., and Athanasopoulos, G.,
Forecasting: Principles and Practice, OTexts, 2018.

%%%%%%%%%%%%%%% REf 22 (dataset) %%%%%%%%%%%%%%%%%%%%
\bibitem{makridakis2018m4}
Makridakis, S., Spiliotis, E., and Assimakopoulos, V.,
The M4 Competition: Results, findings, conclusion and way forward,
International Journal of Forecasting, 2018.


%%%%%%%%%%%%%%% REf 23 (dataset) %%%%%%%%%%%%%%%%%%%%
\bibitem{uci_electricity}
Dua, D., and Graff, C.,
UCI Machine Learning Repository: Electricity Load Diagrams 2011--2014,
2017. [Online]. Available: https://archive.ics.uci.edu/ml/datasets/ElectricityLoadDiagrams20112014

%%%%%%%%%%%%%%% REf 24 (dataset) %%%%%%%%%%%%%%%%%%%%
\bibitem{uci_pgcb}
UCI Machine Learning Repository,
PGCB Hourly Generation Dataset (Bangladesh),
2025. [Online]. 
Available: https://archive.ics.uci.edu/dataset/1175/pgcb+hourly+generation+ dataset+(bangladesh)


%%%%%%%%%%%%%% REF 25 %%%%%%%%%%%%%%%%%%%%%%%%%%%%%
\bibitem{box2015time} 
G. E. P. Box, G. M. Jenkins, G. C. Reinsel, and G. M. Ljung, \emph{Time Series Analysis: Forecasting and Control}, 5th ed. Hoboken, NJ, USA: Wiley, 2015.

%%%%%%%%%%%%%% REF 26 %%%%%%%%%%%%%%%%%%%%%%%%%%%%% 
\bibitem{erickson2020autogluon} 
N. Erickson \emph{et al.}, "AutoGluon-Tabular: Robust and accurate AutoML for structured data," \emph{arXiv preprint arXiv:2003.06505}, Mar. 2020.





\end{thebibliography}
\end{document}